\documentclass[sigconf]{acmart}

\usepackage{multirow}
\usepackage{subcaption}
\usepackage{longtable}
\usepackage{array}
\usepackage[singlelinecheck=off]{caption}
\usepackage[colorinlistoftodos]{todonotes}
\usepackage[title]{appendix}

\AtBeginDocument{%
  \providecommand\BibTeX{{%
    \normalfont B\kern-0.5em{\scshape i\kern-0.25em b}\kern-0.8em\TeX}}}

\setcopyright{none}
\renewcommand\footnotetextcopyrightpermission[1]{} 
\begin{document}

\title{Comparing Deep Learning Models for the Task of Volatility Prediction Using Multivariate Data}



\author{Wenbo Ge}
\email{wenbo.ge@anu.edu.au}
\orcid{0000-0002-0429-3061}
\affiliation{%
  \institution{The Australian National University}
  \department{School of Computing}
  \streetaddress{145 Science Road, Acton}
  \city{Canberra}
  \state{ACT}
  \postcode{2601}
  \country{Australia}}

\author{Pooia Lalbakhsh}
\email{pooia.lalbakhsh@monash.edu}
\orcid{0000-0001-9267-2610}
\affiliation{%
  \institution{Monash University}
  \department{Monash Energy Institute\\Department of Data Science \& Artificial Intelligence}
  \streetaddress{20 Exhibition Walk}
  \city{Clayton}
  \state{Victoria}
  \postcode{3168}
  \country{Australia}}

\author{Leigh Isai}
\email{leigh@eulercapital.com.au}
\orcid{0000-0003-3717-3477}
\affiliation{%
  \institution{Euler Capital}
  \streetaddress{63-67 Cemetery Road}
  \city{Drysdale}
  \state{Victoria}
  \postcode{3222}
  \country{Australia}}

\author{Artem Lensky}
\email{artem@lensky.com.au}
\orcid{0000-0002-4745-6756}
\affiliation{%
  \institution{Lensky Analytics}
  \department{}
  \streetaddress{}
  \city{Braddon}
  \state{ACT}
  \postcode{2601}
  \country{Australia}}

\author{Hanna Suominen}
\email{hanna.suominen@anu.edu.au}
\orcid{0000-0002-4195-1641}
\affiliation{%
  \institution{The Australian National University}
  \department{School of Computing}
  \streetaddress{145 Science Road, Acton}
  \city{Canberra}
  \state{ACT}
  \postcode{2601}
  \country{Australia}}
\affiliation{%
  \institution{University of Turku}
  \department{Department of Computing}
  \city{Turku}
  \postcode{FI-20014}
  \country{Finland}}

\renewcommand{\shortauthors}{Ge et al.}

\begin{abstract}
This study aims to compare multiple deep learning-based forecasters for the task of predicting volatility using multivariate data. The paper evaluates a range of models, starting from simpler and shallower ones and progressing to deeper and more complex architectures. Additionally, the performance of these models is compared against naive predictions and variations of classical GARCH models.

The prediction of volatility for five assets, namely S\&P500, NASDAQ100, gold, silver, and oil, is specifically addressed using GARCH models, Multi-Layer Perceptrons, Recurrent Neural Networks, Temporal Convolutional Networks, and the Temporal Fusion Transformer. In the majority of cases, the Temporal Fusion Transformer, followed by variants of the Temporal Convolutional Network, outperformed classical approaches and shallow networks. These experiments were repeated, and the differences observed between the competing models were found to be statistically significant, thus providing strong encouragement for their practical application.
\end{abstract}


\settopmatter{printfolios=true}
\maketitle

\section{Introduction}

\noindent One of the most important tasks for a financial institution is to monitor the volatility of its portfolio. Volatility is a key factor in security valuation, risk management, and options pricing, as well as it affects investment choices and valuation of public and corporate liabilities \cite{enow2023modelling}. Needless to say, being able to forecast volatility and the many components affected by it would be extremely beneficial. 
    
However, given that price aggregates a myriad of factors including economic fundamentals, market sentiment, market makers, as well as the internal market dynamics affected by the orderbook and stop-losses \cite{Liston_2023}, sophisticated computational models are developed to support practitioners' judgment and decision making in quantitative finance.
The stylised facts of volatility \cite{guillaume_birds_1997,engle_2_2007,masset_volatility_2011}, the efficient market hypothesis \cite{malkiel_efficient_1989, timmermann_efficient_2004}, and the ephemeral nature of financial relationships \cite{clarida_monetary_2000,edwards_stock_2003} are just a few of the reasons why designing such forecasting models, developing their evaluation settings, and deploying the evaluated models in the specifics of the task are complex.

By proceeding from simpler or shallower models (i.e., the GARCH models and \textit{Multi-Layer Perceptrons} (MLP)) to deeper and more complex NNs (i.e., the \textit{Recurrent NNs} (RNN), \emph{Convolutional NNs} (CNN), \textit{Temporal Convolutional Networks} (TCN), and \textit{Temporal Fusion Transformer} (TFT)),  we will forecast the volatility of five assets (i.e., S\&P500, NASDAQ100, gold, silver, and oil). These carefully developed examples with performance evaluations and statistical analyses should encourage practitioners to study and apply deep learning (DL) in forecasting volatility as a way of improving the accuracy of predicted volatility.
%
%
%


\section{Background and literature review} \label{sec: formulating task}

\subsection{Defining volatility}

Volatility is often referred to as latent and unobservable \cite{poon_forecasting_2003}. Consequently, there are many different ways to quantify volatility \cite{patton_volatility_2011}, such as \textit{historical volatility} (HV, the standard deviation of logarithmic returns over a time window) \cite{poon_forecasting_2003},\footnote{Note that despite the name containing the word `historical', it is not defined exclusively for historical data. This can still be forecasted in the same way that realized volatility can forecast. Alternatively, HV is sometimes referred to as the close-to-close method.} \textit{realised volatility} (RV, the square root of the sum of squared log-returns over a time window) \cite{andersen_answering_1998}, \textit{implied volatility} (IV, backwards calculated from options prices via an option pricing model, such as Black-Scholes) \cite{mayhew_implied_1995}, and many more. Each definition has different properties, discussed in more detail in \cite{ge_neural_2022}, and which definition to use should reflect the aspect of volatility that needs to be captured and the purpose of forecasting volatility. If we are trying to forecast options prices, IV is a natural choice, as it is one of the underpinning variables in its pricing. IV can also be good if we are trying to gain insight into the market sentiment of certain assets, as the IV of the S\&P index is commonly referred to as the `fear' and `greed' index \cite{vergili2023relationship}. Alternatively, if we are more interested in the changes in price or returns of a given asset, we may decide to use a HV or RV.

In addition to the definition of volatility, we must also decide on the window of time that this volatility captures, keeping in mind that the further into the future we are trying to forecast, the more uncertain any forecast will be. Again, this should depend on the reason we are forecasting volatility; for example, market makers and day traders may want to monitor short-term volatility in the span of minutes to watch for entry/exit signals.
The volatility in the very long term (such as over the next year) will hold almost no useful information as the time span is so wide, and will also be very hard to forecast as the forecast period is so far from the information period. Volatility in the very short term may also not be useful for the investors, as they do not intend to action anything over the seconds or minutes time frame. However, predicting short-term volatility is of interest to market makers and high-frequency arbitrage traders \cite{Hao_2023}.

\subsection{Motivation}


The efficient market hypothesis is sometimes interpreted to mean that, with a good enough model, the market price of assets is the only data required as it captures all public information \cite{nazlioglu2023efficient}. The reality is that even if market prices were the culmination of all public information, not all public information can be decoded from market prices, the relationship is not injective, and thus not invertible. Despite this, it is often the case that the market price of the asset of interest is enough information for a decent forecast, especially in near and short-term volatility forecasts as trends and patterns can be detected. 


In many other cases, introducing additional data that may capture useful information and that can be readily decoded is helpful. This is not necessarily limited to market prices or even financial data.
For example, information on the economy of a country can be interpreted from the \textit{Gross Domestic Product} (GDP) or its unemployment rate, and the level of inflation can be drawn from the \textit{Consumer Price Index} (CPI), the purchasing power of a currency can be derived from its exchange rates to several highly traded currencies, and the market sentiment may be derived from the options prices for given assets. The time of day or the day of the week may also act as a conditional for some models, as trading patterns may change in different periods.
The public's sentiment about the market and its volatility can be decoded from Tweets and Google Search phrases \cite{kordonis_stock_2016,ren_forecasting_2019}. Information regarding what the public might do, and thus how markets will move, may be held in news articles or the social media of influential individuals; a `follow-the-leader' phenomenon \cite{dittmar_it_2012, zhang_exploiting_2018}. Insider information may be reflected in portfolio changes of well-connected individuals \cite{lakonishok_are_2001,neupane_mimicking_2021}.

The time period for which the data is useful also needs to be considered. In other words, we should consider the time frame of useful information to provide to the model for training and inference. If the goal was to forecast volatility in the short term, such as the next 15 minutes, it may be a waste to use data from the previous 50 years, as the trading strategies and market regimes in the early years are very likely to be different than in the later years. However, using only a week's worth of training data may also miss out on information from past market regimes that may repeat. 
The amount of information provided at inference time is also important as it impacts the computation time, as well as may dilute the useful information. When inferring the volatility of the next 30 days, there is likely to be useful information held the day before, but that alone may not be enough. However, using all the data from the beginning of time will dilute the useful information, as well as cause the model to compute for a long time for inference.  

How the data are framed and presented to the model is also a very important factor. The same data (e.g., the daily closing market prices of several assets) can be presented in many different ways to a model. In its simplest form, a string of numbers, each representing the close price of a given day, can be indexed by date or not. 
Often, this data representation is normalized by using one or more of many available normalization algorithms (e.g., 0-1, logarithmic, unit vector, or min-max) before model training and inference. 
The data could also be transformed into a graphical representation e.g. an image so that the dependencies and explainability are better incorporated into the system \cite{Hao_2023}. 

Finally, the timing of the data must be considered. If using daily prices, some assets are not on the market every day, and as a result, when using multiple data streams, 
a mismatch between the date and time of each point in the time series is likely to be present; a simple cure is to fill in the missing data between points by using interpolation. 
One aspect that should be clarified is that future information should not be fed back for any kind of testing (this would violate good practice; 
Leaving empty days and using a forecasting model that is robust to this provides another alternative. 
To account for all the aforementioned nuances, models of certain structures with a high number of parameters are required. Deep learning models are a good fit for the task and deserve investigation.

\subsection{Forecasting models for financial volatility} \label{sec: forecasting models}

\noindent One of the most widely used family of models is the \textit{auto-regressive} (AR) models, which encompasses the well-known GARCH model \cite{bollerslev_generalized_1986} and its many variants. Each variant captures different properties of volatility that may be more important for each individual task and typically do not include exogenous variables. Intelligent models from machine learning(ML) and DL have also shown much success \cite{cavalcante_computational_2016, ismail_fawaz_deep_2019, chen2022volatility}, rising in popularity due to their successful application in many other fields. However, until recent years the intelligent models used in the \textit{State-Of-The-Art} (SOTA) financial volatility forecasting tend to be much shallower and simpler than those typically used in DL \cite{ge_neural_2022}. 

Of the many types of models that can be used to understand and forecast volatility, none are as widespread as the AR family of models. The seminal \textit{Auto-Regressive Conditional Heteroscedasticity} (ARCH) model, initially proposed in 1982, models future volatility conditioned on previous observations \cite{engle_autoregressive_1982}. This was then adapted to the GARCH model \cite{bollerslev_generalized_1986} by including an \textit{Auto-Regressive Moving Average} (ARMA) component in the model \cite{bollerslev_generalized_1986}. Since then, there have been many advancements that attempt to address the models' inability to capture several stylized facts of volatility \cite{engle_2_2007}. Several examples are the \textit{Exponential}, \textit{Threshold}, and \textit{Glosten-Jagannathan-Runkle} versions (EGARCH, TGARCH, GJR-GARCH) which allow for asymmetric dependencies in volatility, and the \textit{Integrated} and \textit{Fractionally Integrated} versions (IGARCH, FIGARCH) which address volatility persistence, where an observed shock in the volatility series seems to impact future volatility over a long horizon. Despite the many variants of the GARCH model, several experiments have found that the simple ARCH($1$) and GARCH($1$, $1$) forecasting models perform the best \cite{hansen_forecast_2005,orhan_comparison_2012}.

Neural network (NN) based models are commonly used in various areas including volatility prediction \cite{ge_neural_2022}, and although they do not have the same theoretical underpinnings as the aforementioned GARCH models, they are extremely flexible, often referred to as universal function approximators \cite{hornik_multilayer_1989}, possessing the ability to learn any arbitrary mapping $f$ from input $\textbf{X}$ to output $y$; $y = f(\textbf{X})$ \cite{white_connectionist_1990}. In the context of time series analysis, a \textit{Nonlinear Auto-Regressive} (NAR) framework is often adopted with the MLP, enforcing an AR property to the nonlinear mapping (e.g., $\widehat{y}_{t+1} = f([y_t,y_{t-1},...,y_{t-m}]^T)$) \cite{lapedes_nonlinear_1987, chow_neural_1996, kumar_p_estimation_2015}. This can be extended into the framework of the nonlinear autoregressive network with exogenous inputs (NARX) by including such input variables as those derived from several indices, exchange rates, and outputs of GARCH models, thus providing more information to the model \cite{lin_learning_1996, bucci_choleskyann_2020} which has been beneficial for forecasting performance \cite{kristjanpoller_volatility_2017}. Other network architectures, such as the RNN originally designed for sequential data, have also been used in volatility forecasting \cite{hajiabotorabi_improving_2019}. Variations of the RNN, such as \textit{Long Short-Term Memory} (LSTM) models, can also be used, for instance in \cite{kim_forecasting_2018} whereby an LSTM and GARCH model were combined to forecast HV. CNNs and other NNs used in computer vision can also be exploited by first converting the time series into an image;  gold prices can be converted into a 3-channel RGB image and then processed with a pre-trained vgg16 (a well-known and high-performing CNN model \cite{simonyan_very_2015}) \cite{vidal_gold_2020}.

Although RNN, LSTM, and CNN are considered to be deep models, they are not considered SOTA for time-series processing in DL, a distinct gap highlighted in a recent systematic literature review \cite{ge_neural_2022}. This is reserved for recent models that have extremely deep capacity and use complex models, often adapted from other fields such as TCNs, 
which have been successful in music generation, speech enhancement, and many other areas involving time series \cite{oord_wavenet_2016, lea_temporal_2017, pandey_tcnn_2019}. The TCN was an adaptation of the CNN, consisting of 1-dimensional convolutional blocks structured in a way that does not violate the temporal ordering of data (that is, only past data can be seen when forecasting), known as a causal convolution \cite{oord_wavenet_2016}. In conjunction with a progressively increasing dilation size, the receptive field can be exponentially increased as layers increase, thus allowing for the exploitation of long-term relationships. These blocks also often use residual connections, layer normalization, gradient clipping, and dropout, all of which have been shown to improve learning and performance \cite{pascanu_difficulty_2013, ba_layer_2016, he_deep_2016, bai_empirical_2018, zhang_why_2020}. 
 

The Transformer was originally designed for natural language processing and handles sequential data well \cite{vaswani_attention_2017}. It has outperformed several top models in different natural language processing tasks and also performed well in unrelated tasks without being trained for them, leading to the hypothesis that the Transformer is a general model \cite{brown_language_2020}. The key building block in the Transformer is multiheaded self-attention, which allows the data to be encoded in a more meaningful way. The decoder then makes a prediction, conditional on the inputs given. Several variants of the Transformer have been specifically adapted to time series data, such as the TFT. The TFT builds on top of the Transformer by deploying a gating mechanism to skip unused components of the network, variable selection networks to select relevant input variables at each time step, static co-variate encoders to provide context to the model, temporal processing to learn long and short-term relationships, and quantile predictions to forecast with corresponding confidence \cite{lim_temporal_2020}.

\section{Methodology} \label{sec: experiment}

\begin{table}[tb]
\caption{\label{tab: data description} Description of data}
\centering
\begin{tabular}{@{}ccc@{}}
\toprule
 Asset & Start date & End date \\
\midrule
 S\&P500 & 22/Sep/2003 & 31/Dec/2018 \\
 NASDAQ100 & 01/Jan/2003 & 31/Dec/2018 \\
 Gold & 03/Jun/2008 & 31/Dec/2018 \\
 Silver & 16/Mar/2011 & 31/Dec/2018 \\
 WTI Crude Oil & 10/May/2007 & 31/Dec/2018 \\
\bottomrule
\end{tabular}
\end{table}

\subsection{Data}

\subsubsection{Main data}
Volatility was forecast for five assets: S\&P500, NASDAQ-100 (NDX), gold, silver, and oil. The data for each, as well as the corresponding volatility indices, was collected  (Table~\ref{tab: data description}). The data consisted of the daily closing prices, as well as the open, high, and low prices for S\&P500, NDX, gold, and oil. Volume was available only for S\&P500 and NDX. Each asset was restricted to a starting date that corresponded with when the volatility index was introduced, except for S\&P500 and NDX. This was because the volatility index for S\&P500 was originally for the S\&P100 and later changed on 22 September 2003 and because the volatility index for NDX began earlier than one of the exogenous variables used. Additionally, the ending date was restricted to 31 December 2018.

\subsubsection{Exogenous/extra data}
\begin{figure}[b]
  \centering
  \includegraphics[trim={0cm 26.5cm 3.1cm 0cm},clip, width=0.9\columnwidth]{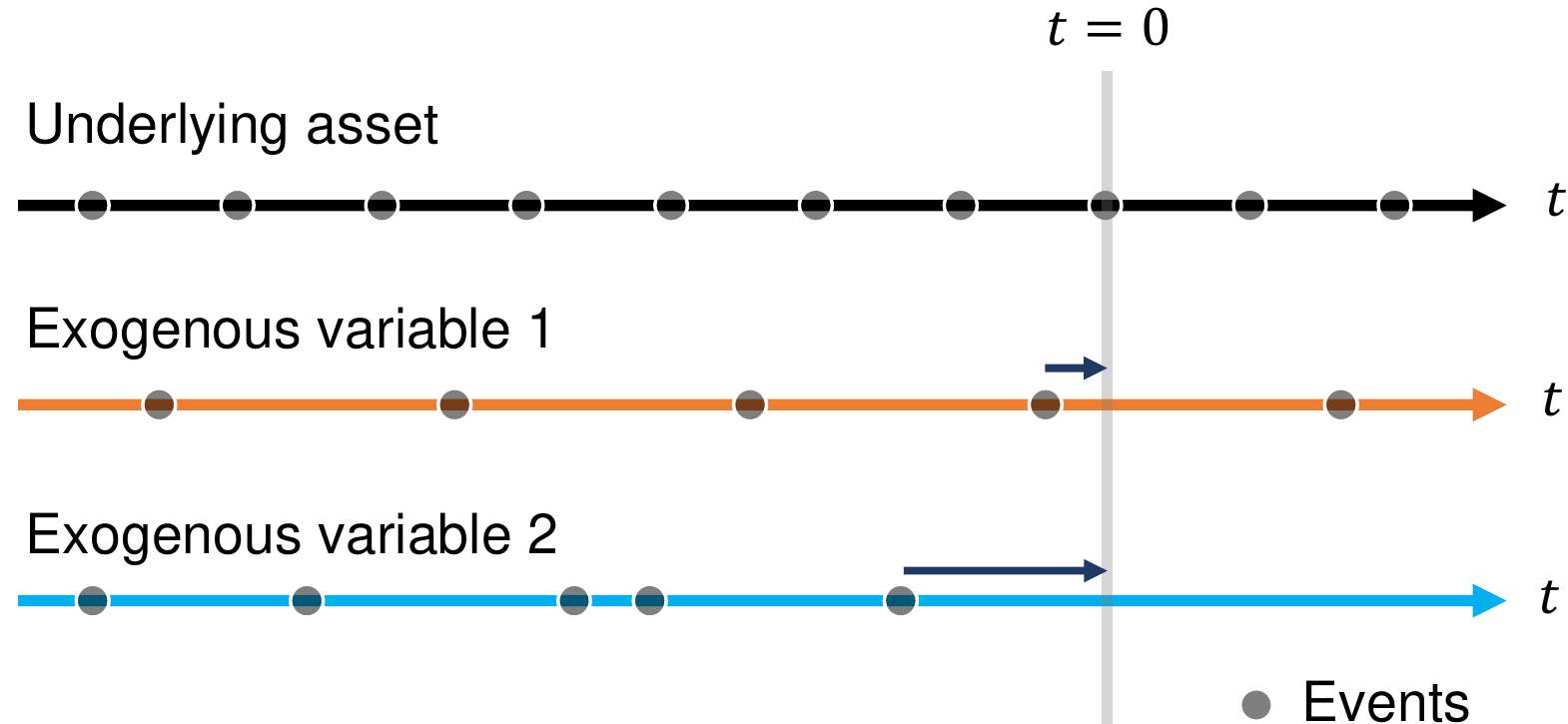}
  \caption{Date matching method for exogenous variables}
  \label{fig: exogenous date match}
\end{figure}

Exogenous variables consisted of indices such as SZSE, BSE SENSEX, FTSE100 and DJIA, exchange rates US-YEN, US-EURO and the US dollar trade-weighted index as well as such United States fundamentals as Federal Reserve primary credit rate, the mean and median duration of unemployment, consumer price index inflation rate, government debt per GDP, gross Federal debt, and currency in circulation. Table \ref{talbe: model inputs} lists all types of input variables used per model. All variables were date matched with the underlying assets by bringing forward the nearest historical value, as depicted in Fig. \ref{fig: exogenous date match}.

\begin{table*}[ht]
\caption{\label{talbe: model inputs} Additional input variables for forecasting models}
\centering
\begin{tabular}{@{} r m{40em}@{}}

\toprule
 Model & Input \\
\midrule
 NARX & Log returns, squared log returns, inverse price of the underlying asset, na\"ive forecasts for HV, na\"ive forecasts for IV, US dollar trade-weighted index, primary credit rate, and median length of unemployment\\
 
 \addlinespace
 \addlinespace
 
 ANN-GARCH & Log returns of the indices: [BSE SENSEX, FTSE100, SZSE, and DJIA], log-returns of the exchange rates: [US-YEN and US-EURO], the na\"ive forecasts for HV, the na\"ive forecasts for IV, and the log-returns of the underlying asset. \\
 
 \addlinespace
 \addlinespace
 
 TCN & Log returns of underlying variable, na\"ive forecasts for HV, na\"ive forecasts for IV, the current direction of movement, US dollar trade-weighted index, Federal Reserve primary credit rate, mean duration of unemployment, the median duration of unemployment, consumer price index inflation rate, Government debt per GDP, gross Federal debt, and currency in circulation \\
 
 \addlinespace
 \addlinespace
 
 TFT & Open, high, low, close, volume (where possible), log returns, squared log returns, inverse price of the underlying asset, na\"ive forecasts for HV, na\"ive forecasts for IV, the US dollar trade-weighted index, Federal Reserve primary credit rate, mean duration of unemployment, the median duration of unemployment, consumer price index inflation rate, Government debt per GDP, gross Federal debt, currency in circulation, day of the week, month of the year, and number of days since previous observation \\

\bottomrule

\end{tabular}
\end{table*}

\begin{figure}[t!]
\captionsetup{justification=centering}
\subcaptionbox{HV}
{\includegraphics[width=\columnwidth,trim={0cm 0cm -1cm 0cm}, clip]{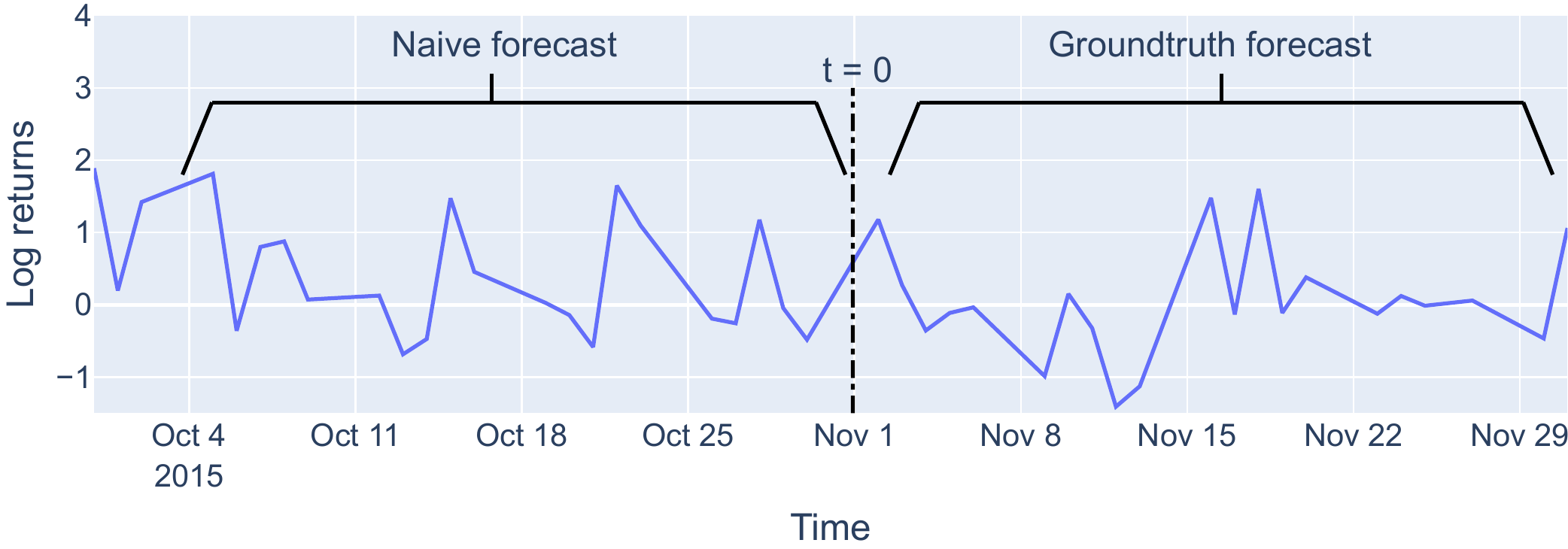}}

\captionsetup{justification=centering}
\subcaptionbox{IV}
{\includegraphics[width=\columnwidth,trim={0cm 0cm -1cm 0cm}, clip]{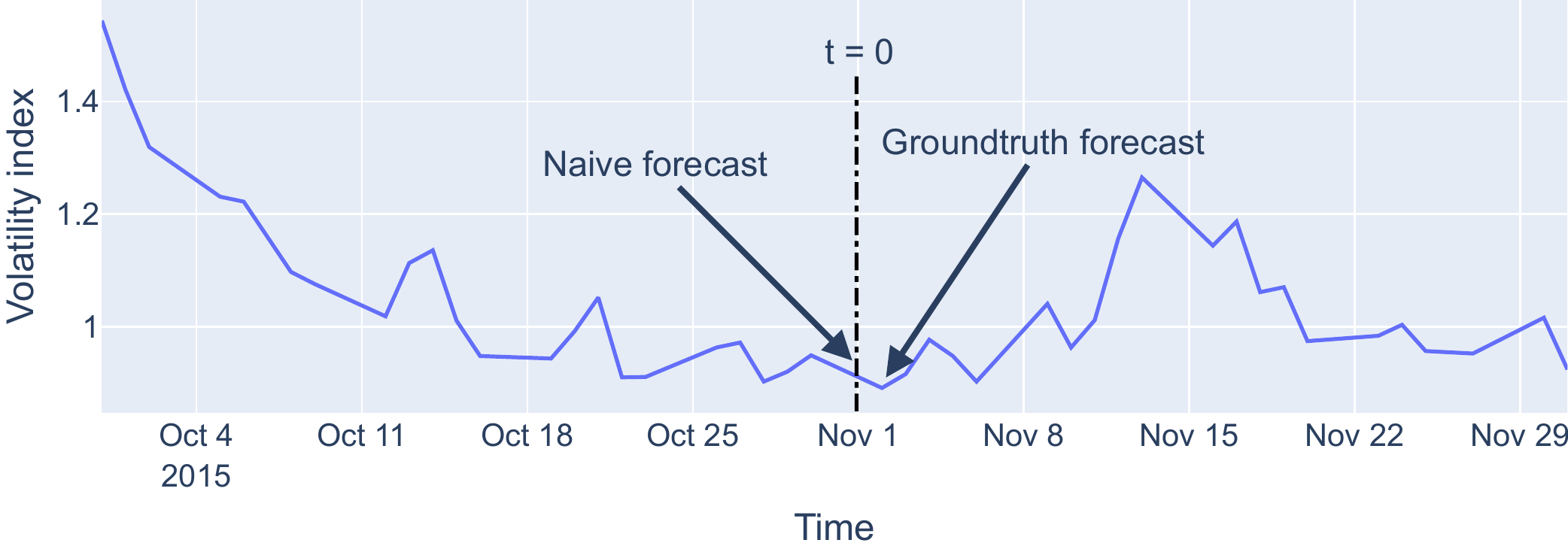}}
\caption{Groundtruth and na\"ive forecasts.}
\label{fig: groundtruth and naive}  
\end{figure}

The task was to forecast the volatility of the next month for the five assets; more specifically, to forecast the month-long HV and IV (Fig. \ref{fig: groundtruth and naive}), starting from 1 day ahead, for S\&P500, NDX, gold, silver, and oil. The ground truth for HV was the standard deviation of the logarithmic returns from 1 trading day ahead to 22 trading days ahead (which is equivalent to one calendar month) \cite{poon_forecasting_2003}. In other words, with $t$ referring to the current time,
we defined over a certain period 
\begin{equation*}
[\tau_1, \, \tau_2 ] = [t+1, \, t+22]
\end{equation*}

\noindent HV as the standard deviation ($\text{std}(\cdot)$) of log-returns as follows: 
\begin{equation} \label{HV equation}
   \text{HV} = \sqrt{\frac{1}{N} \sum_{t=\tau_1}^{\tau_2}{\left(r_t - \frac{1}{N} \sum_{t=\tau_1}^{\tau_2}{r_t} \right)^2}} = 
    \text{std} \left( 
    \begin{bmatrix}
           r_{\tau_2} \\
           r_{\tau_2 - 1} \\
           r_{\tau_2 - 2} \\
           \vdots \\
           r_{\tau_1}
         \end{bmatrix}
         \right) 
\end{equation}
where
$N = \tau_2 - \tau_1 = 21$ 
is the number of samples between the time steps,
$P_t$ is price at time $t$, and
$r_t = \log(P_t/P_{t-1}) \cdot 100$. 

For IV \cite{mayhew_implied_1995}, this meant that the ground truth was simply the value of the volatility index for the next trading day, as the volatility index was already defined for the next calendar month. The values of the volatility indices were also adjusted by a factor of $1/\sqrt{252}$, deannualizing the value to be on the same scale as HV. 

\subsection{Methods}

\begin{figure*}[tb!]
  \centering
\includegraphics[trim={0cm 16.3cm 2.7cm 0cm},clip,width=0.8\textwidth]{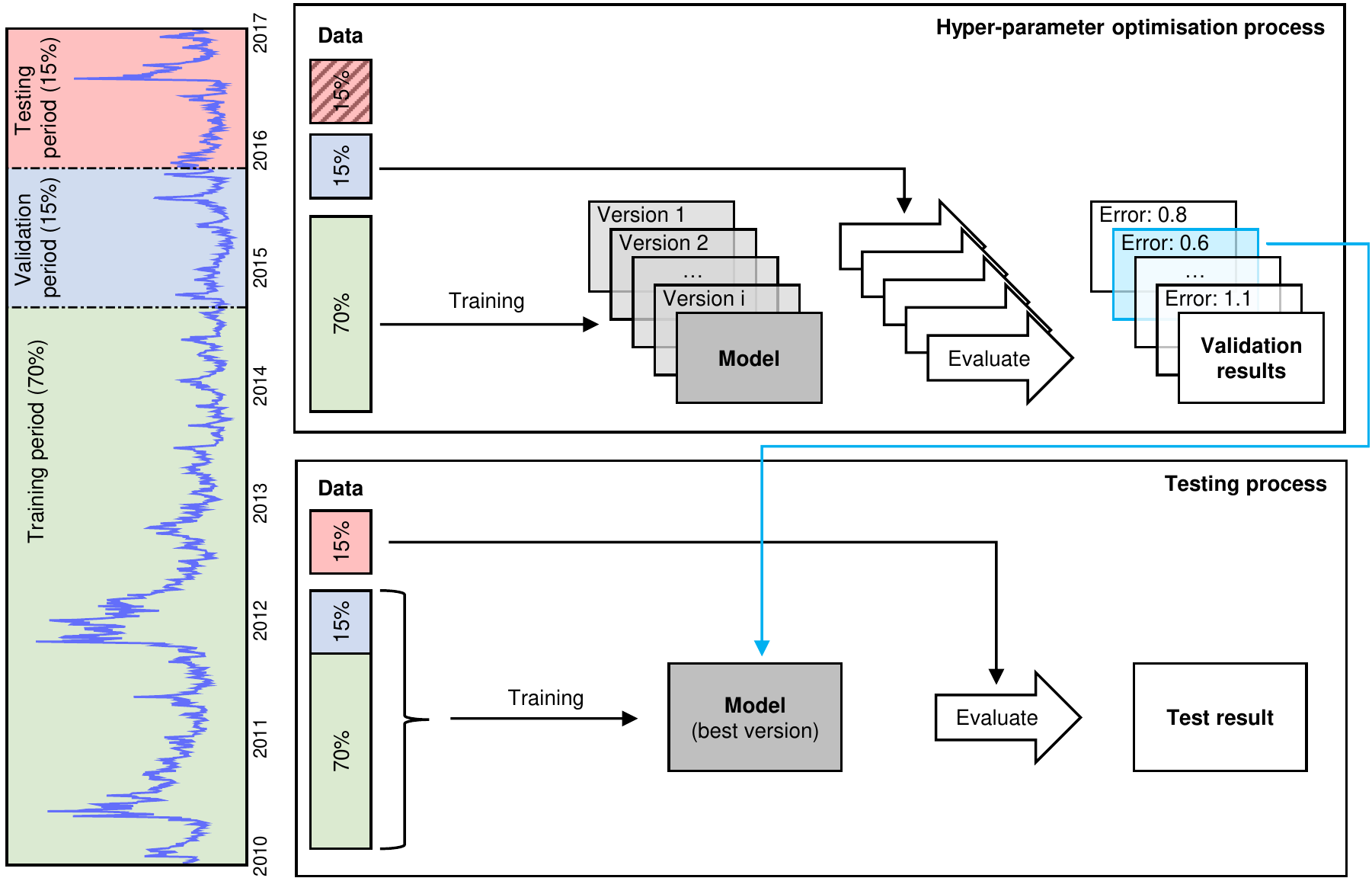}
  \caption{Depiction of a 70-15-15 train-validation-test split with the evaluation process}
  \label{fig: train validation test process}
\end{figure*}

To engineer and evaluate the forecasting models using these five methods, a 70-15-15 train-validation-test split of the data was used \cite{bishop_pattern_2006}  because it did not violate the temporal aspect of the data (Fig. \ref{fig: train validation test process}).
%
The training set was used to train the models using different hyperparameters, which were then evaluated against the validation set to determine the performance with that given set of hyperparameters. Different combinations of hyperparameters were searched (see Appendix \ref{appendix: gridsearch and inputs} for details), and the best-performing set of hyperparameters proceeded to the test phase. Here, the model was re-initialized and trained again with the union of both the training and validation set, then evaluated once using the test set. 
All performances were quantified with the Mean Squared Error (MSE) \cite{bishop_pattern_2006}.

\subsubsection{Statistical significance testing}

Statistical significance testing was also performed in order to distinguish if competing models were statistically significantly different from each other. After the hyperparameters of a model were chosen and the testing phase was completed, the model was reinitialized with a random seed, re-trained, and re-tested. This was repeated until ten MSE values were obtained for each model. These values were then tested across different models in a pair-wise fashion to determine if they were from the same distribution. If so, the models were assumed to be equivalent, if not, then the model with the smaller mean MSE was assumed to be superior. The \textit{Shapiro-Wilk} (SW) test \cite{SHAPIRO1965} was first applied to assess the normality of the distribution with significance level $\alpha$ = 0.05. The choice of the SW test was justified by the applicability of the test to the data with unspecified mean and variance, as well as its appropriateness for small sample sizes. If both distributions were normal then Student's t-test \cite{student1908probable} was used, otherwise the \textit{Kruskal-Wallis} (KW) test \cite{kruskal_use_1952} was employed.

\subsubsection{Description of traditional benchmark models}

The na\"ive model simply repeated the most recent known value of volatility (Fig. \ref{fig: groundtruth and naive}). For IV, this was the value of the corresponding volatility index at the current time. For HV, this was the standard deviation of log returns for the current day and previous 20 days, that is, Equation (\ref{HV equation}) with $\tau_1 = t-20$ and $\tau_2 = t$. This assumed that at the time of forecasting, the current trading day is over and observed, an assumption maintained for all forecasting models. 

Several GARCH variants were reviewed for the best-performing model, along with the lag order of those models.
Namely, the variants consisted of ARCH, GARCH, EGARCH, GJR-GARCH, and GARCH with a Student's t-distribution. 
Lag orders were 1 and 2, as informed by the auto-correlation and partial auto-correlation functions.

Variants of the MLP were also searched over, using both a NAR and NARX framework. For the NAR models, the number of lagged inputs, number of hidden layers, number of neurons, activation function, learning rate, and preprocessing method was tuned. Similarly, the NARX framework searched over the same hyperparameters, with the addition of which exogenous variables to include: the log returns, squared log returns, the inverse price of the underlying asset, the na\"ive forecasts for HV and IV, the primary credit rate, the median length of unemployment, and the US dollar trade-weighted index. 

\subsubsection{Description of benchmark models used for comparison}

The first model picked up from the literature was the ANN-GARCH model \cite{kristjanpoller_volatility_2017}, a combination of the previously mentioned GARCH and NARX MLP models. The GARCH model made a next-day volatility forecast, which is used as an input for the NARX model, along with several other variables that describe the asset and the market. Both the GARCH and NARX models were trained on the prior 252 trading days (one calendar year), and re-trained for every forecast, resulting in a rolling model. The variant and lag order of the GARCH model was searched over as hyperparameters, as well as a number of lagged inputs, the number of hidden layers and neurons, activation functions, preprocessing method, and the set of additional input variables for the NARX model. The possible input variables to include were: the log returns of several indices (BSE SENSEX, FTSE100, SZSE, and DJIA), the log returns of several exchange rates (US-YEN and US-EURO), the na\"ive forecasts for HV and IV, as well as the log-returns of the underlying asset. 

\begin{figure*}[tb!]
  \centering
  \includegraphics[trim={5.5cm 8.5cm 5.8cm 7.9cm},clip,width=0.8\textwidth]{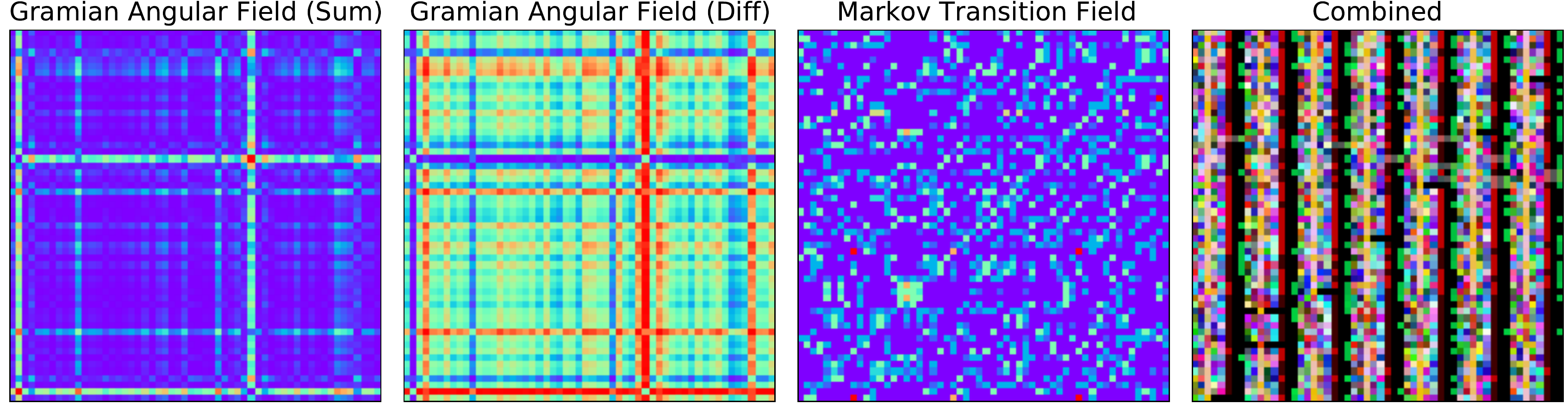}
  \caption{Images generated from Gramian Angular Fields and Markov Transition Fields with 80 trading days of log returns}
  \label{fig: vidal image}
\end{figure*}

The second model was the CNN-LSTM \cite{vidal_gold_2020}. This used the log-returns of the underlying asset, converted to a 3-channel RGB image through Gramian Angular Fields and Markov Transition Fields \cite{campanharo_duality_2011, wang_encoding_2015} (Fig. \ref{fig: vidal image}). Features from these images were extracted using vgg16 \cite{simonyan_very_2015}, and then further processed by several convolutional blocks. In addition to this, the log-returns of the underlying asset were also processed by the LSTM, which was then combined with the processed image features to forecast volatility. The hyperparameters search was performed over the range of learning rates, preprocessing methods, and the length of the sequence that the LSTM was exposed to.

\subsubsection{Experimental models}

Several variations of two deep learning models that recently have been used in predicting volatility are further investigated.

\begin{figure}[tb!]
  \includegraphics[width=0.495\textwidth]{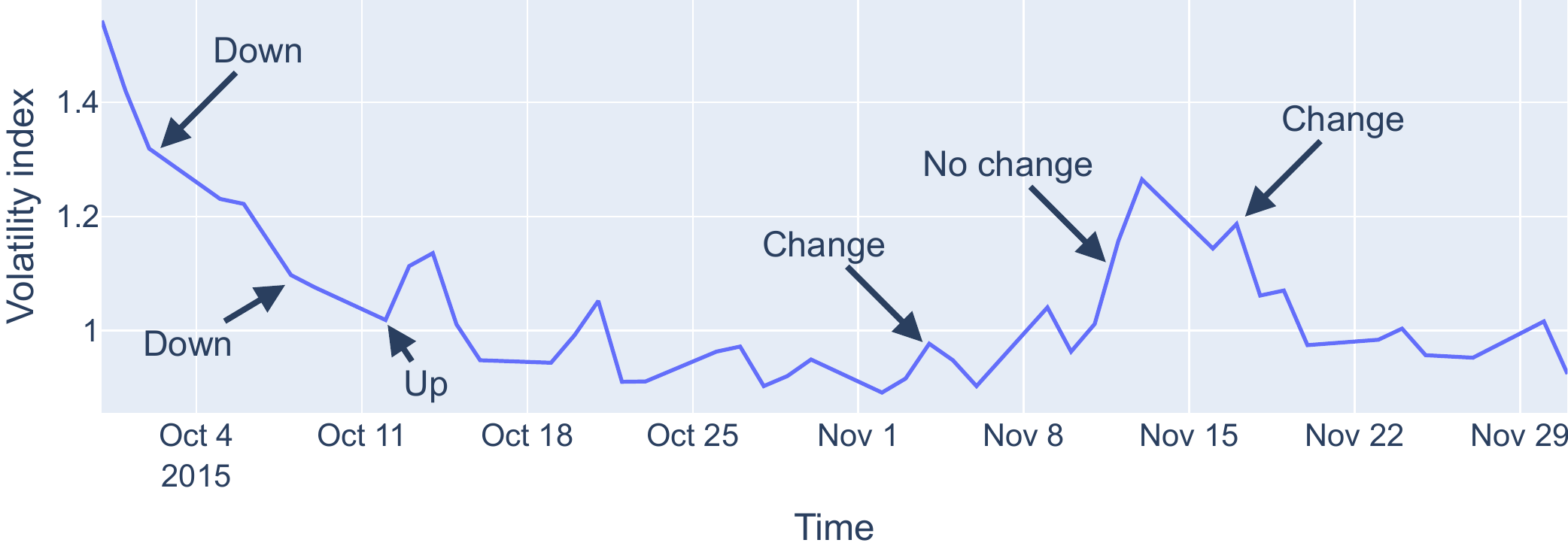}
  \caption{Examples of direction and direction change}
  \label{fig: direction and direction change}
\end{figure}

The first model was the vanilla Temporal Convolutional Network(TCN), as well as the TCN with several modifications. The first modification was to leverage the na\"ive model and forecast a residual, defined as either the difference $y_{\text{groundtruth}, t} - y_{\text{na\"ive}, t}$, or the log difference $log(y_{\text{groundtruth}, t}/y_{\text{na\"ive}, t})$. Another modification was to include multiple tasks to the network, introducing another loss function that will have a separate but related training effect \cite{ruder_overview_2016}. The additional task was to predict either the direction of the forecast (up or down), or the change in direction (change or no change), depicted in Fig. \ref{fig: direction and direction change}. The final modification was to include additional input channels, introducing new information to the network \cite{wan_multivariate_2019}, such as descriptors of the underlying asset (log returns, na\"ive forecasts for HV and IV, and current direction of movement), and variables that describe the market (US dollar trade-weighted index, Federal Reserve primary credit rate, mean and median duration of unemployment, consumer price index inflation rate, Government debt per GDP, gross Federal debt, and currency in circulation) as there is literature to suggest that this may improve performance \cite{kristjanpoller_volatility_2017}. The hyperparameters searched over were as follows: which definition of the residual to use, the form of the additional task, the weighting of the additional task, and the set of variables to include, as well as the number of hidden layers, number of filters, kernel size, dropout, gradient clipping threshold, learning rate, and preprocessing method.

The second model was the Temporal Fusion Transformer (TFT), an adaptation of the Transformer model. Additional variables such as open, high, low, close, and volume were investigated, as well as log returns, squared log returns, inverse price of the underlying asset, and na\"ive forecasts for both HV and IV. Fundamentals such as the US dollar trade-weighted index, Federal Reserve primary credit rate, the mean and median duration of unemployment, consumer price index inflation rate, Government debt per GDP, gross Federal debt, and currency in circulation we also considered, finally such time descriptors as day of the week, month, and a number of days since previous observation, were also included. The hyperparameter search explored the set of input variables, as well as the gradient clipping threshold, hidden size, hidden continuous size, number of attention heads, learning rate, dropout, encoder length, and prediction length. The full details of the hyperparameter search for these models were reported in Appendix \ref{appendix: gridsearch and inputs}.

\subsection{Result evaluation and analysis}
\label{results}

\subsubsection{Benchmark models}

The ANN-GARCH model performed best overall in forecasting HV, achieving the smallest error for three of the assets, as seen in Table \ref{table: benchmark models}. There was no clear GARCH variant that was best among the ANN-GARCH models, however, a lag order of 1 was preferred over all assets. Similarly, there was no dominant MLP structure among  the ANN-GARCH models, every specific task had its own best performing variant. The CNN-LSTM and the MLP models also performed well, achieving the smallest errors for silver and oil, respectively. The CNN-LSTM model preferred the LSTM to be exposed to 42 days for all assets, except S\&P500 for which the model preferred 21 days. The MLP model did not have a consistently preferred lag across the assets but benefited from the inclusion of exogenous variables. For gold and oil, the best exogenous variables were those derived from the underlying asset: inverse price, squared log returns, and na\"ive forecasts of HV and IV. For S\&P500, NDX, and silver, the best exogenous variables were the market variables: risk-free rate, unemployment, and the US dollar index. The na\"ive and GARCH models performed the worst overall, being beaten by all other models. Of the GARCH model, a GARCH(1,1) model consistently outperformed all other variants and lag orders.

For IV forecasting, the na\"ive model performed the best overall, achieving the smallest error for all five assets (Table \ref{table: benchmark models}). Following this, the best models were the CNN-LSTM and MLP. The optimal exposure for the LSTM in the CNN-LSTM was 1 day, presenting its best performance for all assets. The MLP model showed no clear optimal lag order, and the addition of variables did not seem to help with forecasting the IV of gold, silver, and oil. However, including the inverse price, squared log returns, and na\"ive forecasts resulted in better performance for S\&P500 and NDX. There was no clear optimal GARCH variant in both the GARCH model and the ANN-GARCH model, but a lag order of 1 was optimal for all assets. The MLP side of the ANN-GARCH model preferred a lag order of 3, and the inclusion of exogenous variables did not improve performance.

\begin{table}[bt]
\caption{\label{table: benchmark models} Performance (MSE) of benchmark models, forecasting on the test set. Smaller is better, bold is the best.}
\centering
\begin{tabular}{p{0.5cm} p{1.9cm} p{0.75cm} p{0.75cm} p{0.75cm} p{0.75cm} p{0.75cm}}
\toprule
 \textbf{Task} & \textbf{Method} & \textbf{S\&P500} &\textbf{NDX} & \textbf{Gold} & \textbf{Silver} & \textbf{Oil} \\
\midrule
 \multirow{5}{2em}{HV} & Na\"ive &   0.1281  &   0.1887  &   0.0216  &   0.0745  &   0.2799  \\
 & GARCH &   0.2662  &   0.9430  &   0.0191  &   0.0611  &   7.9448  \\
 & MLP   &   0.2935  &   0.3542  &   0.0146  &   0.0666  &   \textbf{0.2080}  \\
 & ANN-GARCH &   \textbf{0.1136}  &   \textbf{0.1735}  &   \textbf{0.0105}  &   0.0563  &   0.2191  \\
 & CNN-LSTM  &   0.1550  &   0.1869  &   0.0554  &   \textbf{0.0488}  &   0.3870  \\
\midrule
 \multirow{5}{2em}{IV} &  Na\"ive &   \textbf{0.0098}  &   \textbf{0.0084}  &   \textbf{0.0013}  &   \textbf{0.0023}  &   \textbf{0.0139}  \\
 & GARCH &   0.2296  &   0.5859  &   0.0263  &   0.0279  &   7.6927  \\
 & MLP   &   0.0172  &   0.0151  &   0.0021  &   0.0032  &   0.0278  \\
 & ANN-GARCH &   0.0248  &   0.0159  &   0.0022  &   0.0036  &   0.0312  \\
 & CNN-LSTM  &   0.0141  &   0.0104  &   0.0064  &   0.0104  &   0.0246  \\
\bottomrule
\end{tabular}
\end{table}

\subsubsection{Hyperparameter optimization}

In testing two hyperparameter optimization methods, BOHB only produced a better forecasting model for the HV of oil, and the IV of gold and silver, whilst grid search resulted in a model with better performance for all other assets (Table \ref{table: gridsearch vs BOHB}). 
Both methods were given roughly the same wall time and were both tested using the un-modified TCN. Though it is difficult to say if one method is superior to the other, the continued use of grid search is justified and was the primary hyperparameter optimization method for the remaining experimental TCN models.

\begin{table}[t]
\caption{\label{table: gridsearch vs BOHB} Performance (MSE) of hyperparameter search for TCN, forecasting on the test set. Smaller is better, bold is the best.}
\begin{tabular}{p{0.5cm} p{1.8cm} p{0.75cm} p{0.75cm} p{0.75cm} p{0.75cm} p{0.75cm}}
\toprule
 \textbf{Task} & \textbf{Method} & \textbf{S\&P500} &\textbf{NDX} & \textbf{Gold} & \textbf{Silver} & \textbf{Oil} \\
\midrule
 \multirow{2}{2em}{HV} & Grid search &   \textbf{0.1467}  &   \textbf{0.1912}  &   \textbf{0.0331}  &   \textbf{0.0364}  &   0.2614  \\
 & BOHB &   0.1591  &   0.2010  &   0.0616  &   0.0800  &   \textbf{0.2107}  \\
\midrule
 \multirow{2}{2em}{IV} & Grid search &   \textbf{0.0098}  &   \textbf{0.0083}  &   0.0029  &   0.0032  &   \textbf{0.0136}  \\
 & BOHB &   0.0126  &   0.0093  &   \textbf{0.0017}  &   \textbf{0.0025}  &   0.0139  \\
\bottomrule
\end{tabular}
\end{table}

\setlength{\tabcolsep}{5pt}

\begin{table}[b]
\caption{\label{table: experimental models} Performance (MSE) of experimental models, forecasting on test set. Smaller is better, bold is the best.}
\begin{tabular}{p{0.5cm} p{2.1cm} p{0.75cm} p{0.75cm} p{0.75cm} p{0.75cm} p{0.75cm}}

\toprule
 \textbf{Task} & \textbf{Method} & \textbf{S\&P500} &\textbf{NDX} & \textbf{Gold} & \textbf{Silver} & \textbf{Oil} \\
\midrule
 \multirow{5}{2em}{HV} &  Na\"ive &   0.1281  &   0.1887  &   0.0216  &   0.0745  &   0.2799  \\
 & Best benchmark &   0.1136  &   0.1735  &   \textbf{0.0105}  &   0.0488  &   0.2080  \\
 & TCN &   0.1467  &   0.1912  &   0.0331  &   0.0364  &   0.2614  \\
 & TCN variants   &   0.1300  &   0.1728  &   0.0153  &   \textbf{0.0315}  &   0.2050  \\
 & TFT &   \textbf{0.0294}  &   \textbf{0.0513}  &   0.0116  &   0.0341  &   \textbf{0.0864}  \\
\midrule
 \multirow{5}{2em}{IV} & Na\"ive &   0.0098  &   0.0084  &   0.0013  &   0.0023  &   0.0139  \\
 & Best benchmark &   0.0141  &   0.0104  &   0.0021  &   0.0032  &   0.0246  \\
 & TCN &   0.0098  &   0.0083  &   0.0029  &   0.0032  &   0.0136  \\
 & TCN variants   &   \textbf{0.0093}  &   \textbf{0.0081}  &   \textbf{0.0013}  &   \textbf{0.0022}  &   \textbf{0.0127}  \\
 & TFT &   0.0094  &   0.0083  &   0.0013  &   0.0025  &   0.0134  \\
\bottomrule
\end{tabular}
\end{table}

\subsubsection{Experimental models}

The TFT model performed best overall for HV forecasting, achieving the smallest errors for S\&P500, NDX, and oil (Table \ref{table: experimental models}). An encoder length of 21 days was optimal for all assets, with no set of input variables that were consistently best. S\&P500 and NDX performed best with the addition of variables that describe time and the underlying asset, gold and silver performed best with the addition of variables that describe time, and oil performed best with the addition of market and time descriptors. TCN variants also performed well, achieving the smallest error for silver when utilizing the secondary task of predicting direction change. There was no consistent variant that performed best: S\&P500 and NDX preferred forecasting the residuals, whilst gold, silver, and oil performed better when utilizing a secondary task. The inclusion of exogenous variables only increased the performance for forecasting gold HV. Additionally, TCN variants consistently performed better than the un-modified TCN (Table \ref{table: experimental models}). The smallest error for gold was achieved by a benchmark model, specifically the ANN-GARCH. 

The TCN variants were the best performing model for IV forecasting, achieving the smallest errors for all assets (Table \ref{table: experimental models}). The optimal modification was to use a secondary task of predicting the direction, as well as forecasting the residuals, consistent amongst all assets. S\&P500, NDX, and gold also benefited from the inclusion of the volatility index value and previous direction of movements. For the TFT model, an encoder length of 10 days was preferred for all assets, except for S\&P500 which preferred a length of 126. There was no consistent set of input variables that performed the best. Silver preferred the addition of variables that describe time, gold preferred the addition of variables that describe the underlying asset and time, whilst S\&P500, NDX, and oil performed best with the addition of all variables available: descriptors of the market, the underlying asset, and time.

\subsubsection{Statistical significance}

\begin{figure*}
\centering
\captionsetup{justification=centering} 
\subcaptionbox{SPX\label{fig:spxHV}}
{\includegraphics[width=0.32\textwidth, trim={3.25cm 7.25cm 4.5cm 7cm}, clip]{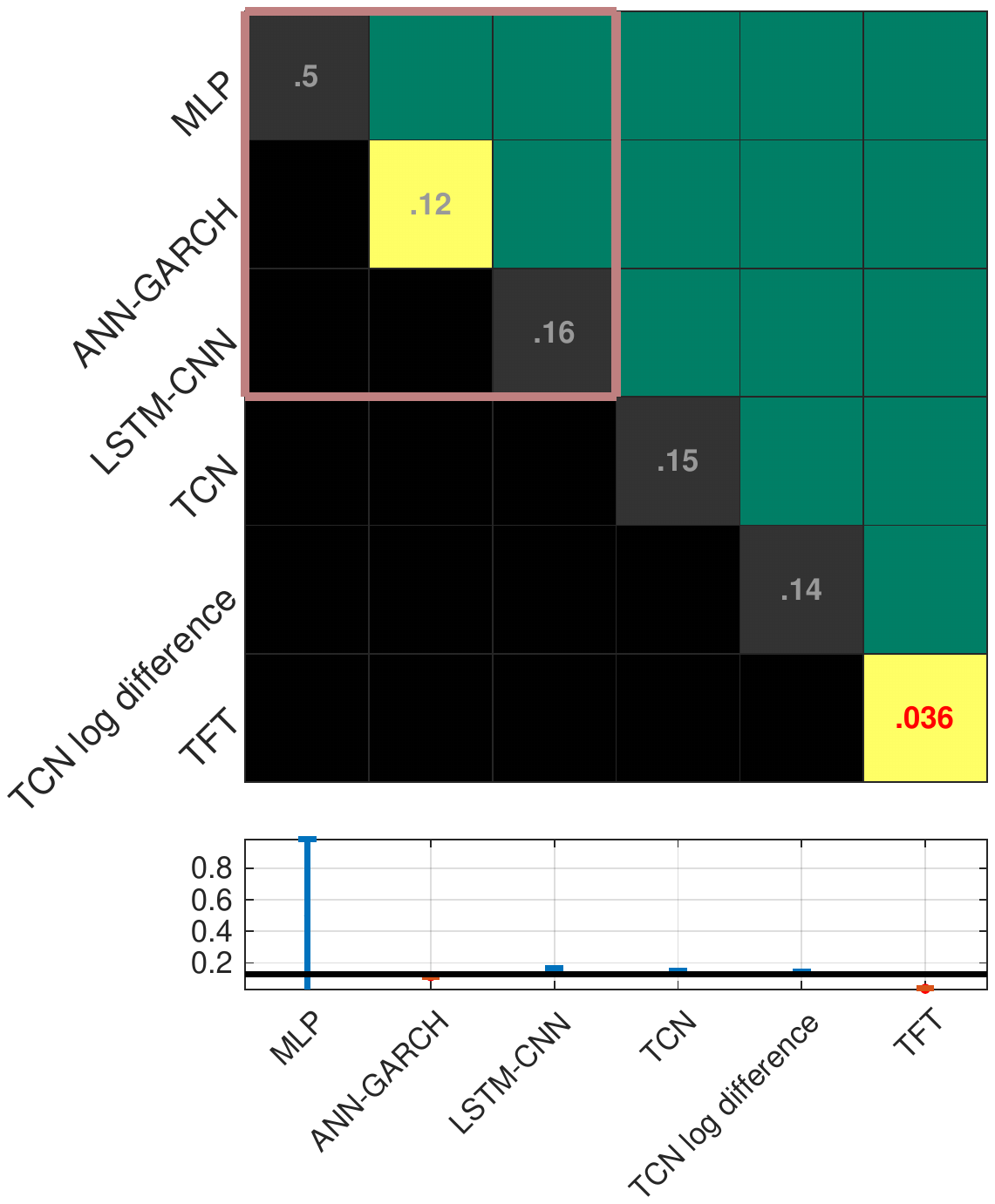}}
\subcaptionbox{Oil}
{\includegraphics[width=0.32\textwidth, trim={3.25cm 7.25cm 4.5cm 7cm}, clip]{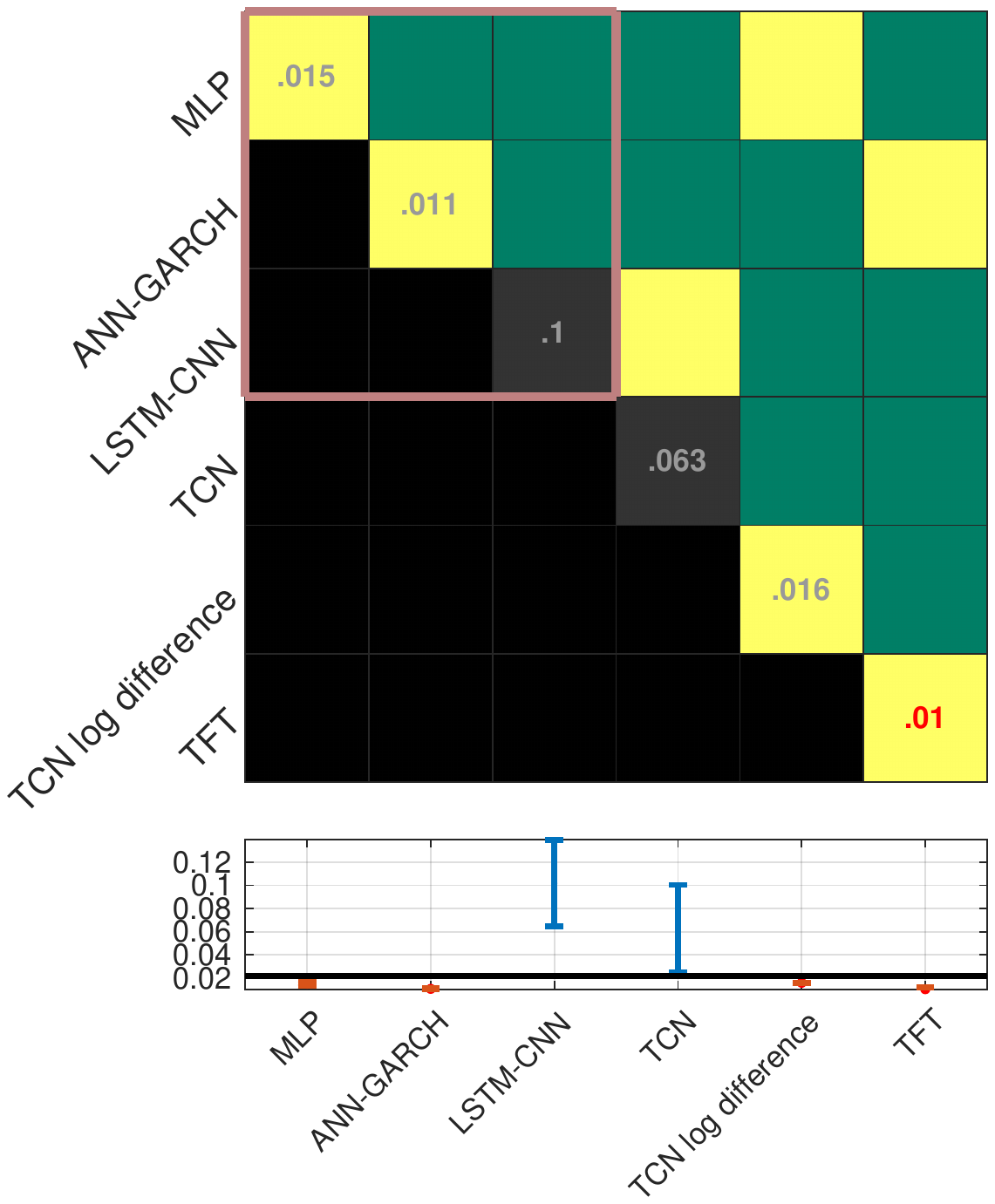}}
\subcaptionbox{Gold}
{\includegraphics[width=0.32\textwidth, trim={3.25cm 7.25cm 4.5cm 7cm}, clip]{Images/goldHV_.pdf}}
\\
\subcaptionbox{Silver}
{\includegraphics[width=0.32\textwidth, trim={3.25cm 7.25cm 4.5cm 7cm}, clip]{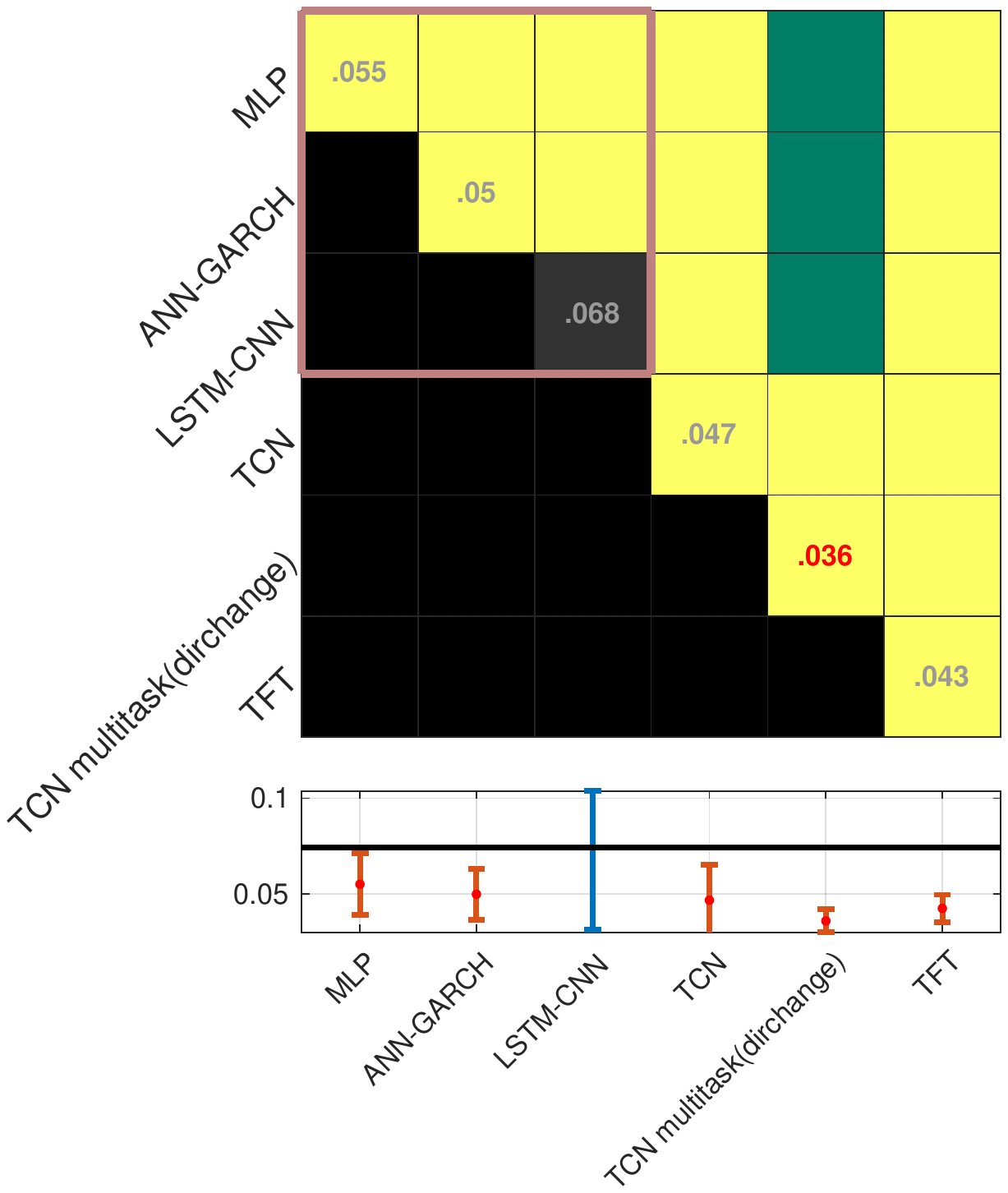}}
\subcaptionbox{Nasdaq}
{\includegraphics[width=0.32\textwidth, trim={2.25cm 6.25cm 4.5cm 7cm}, clip]{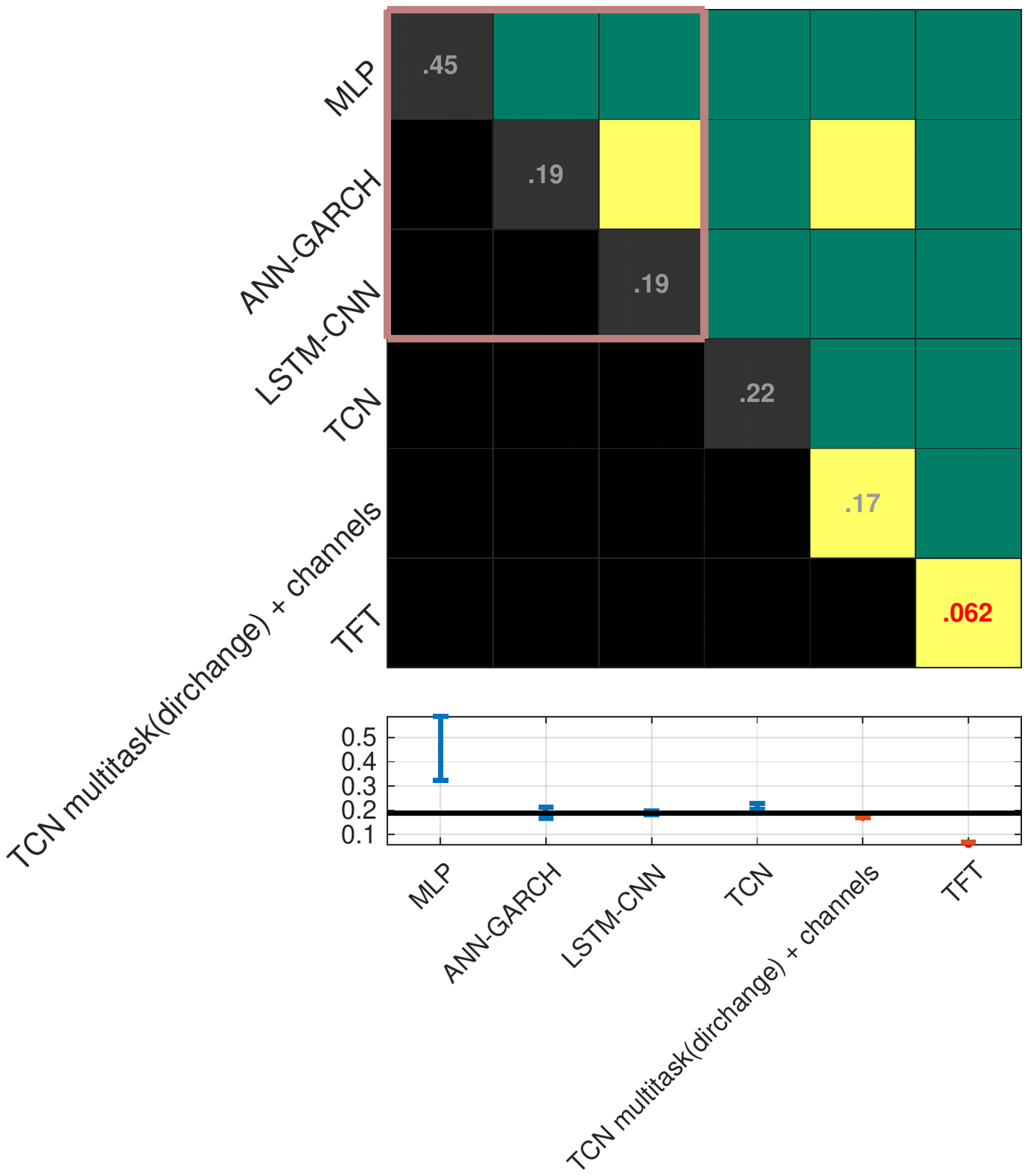}}

\caption{Model comparison of historical volatility prediction using MSE obtained for ten random seed values.}\label{fig:statistical_comparison_HV}
\end{figure*}

\begin{figure*}
\centering
\captionsetup{justification=centering} 
\subcaptionbox{SPX\label{fig:spxHV}}
{\includegraphics[width=0.32\textwidth, trim={3.25cm 7.25cm 4.5cm 7cm}, clip]{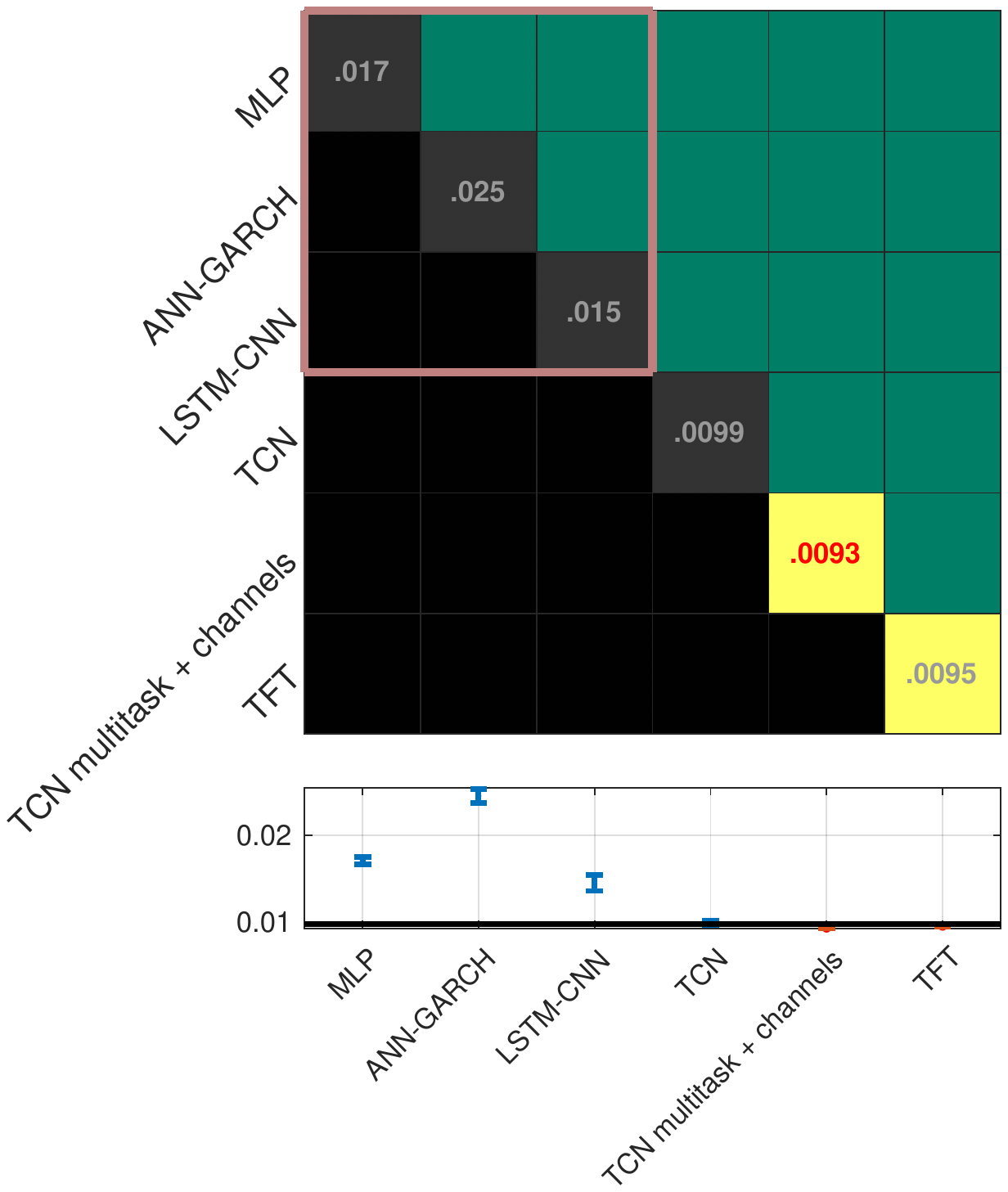}}
\subcaptionbox{Oil}
{\includegraphics[width=0.32\textwidth, trim={3.25cm 7.25cm 4.5cm 7cm}, clip]{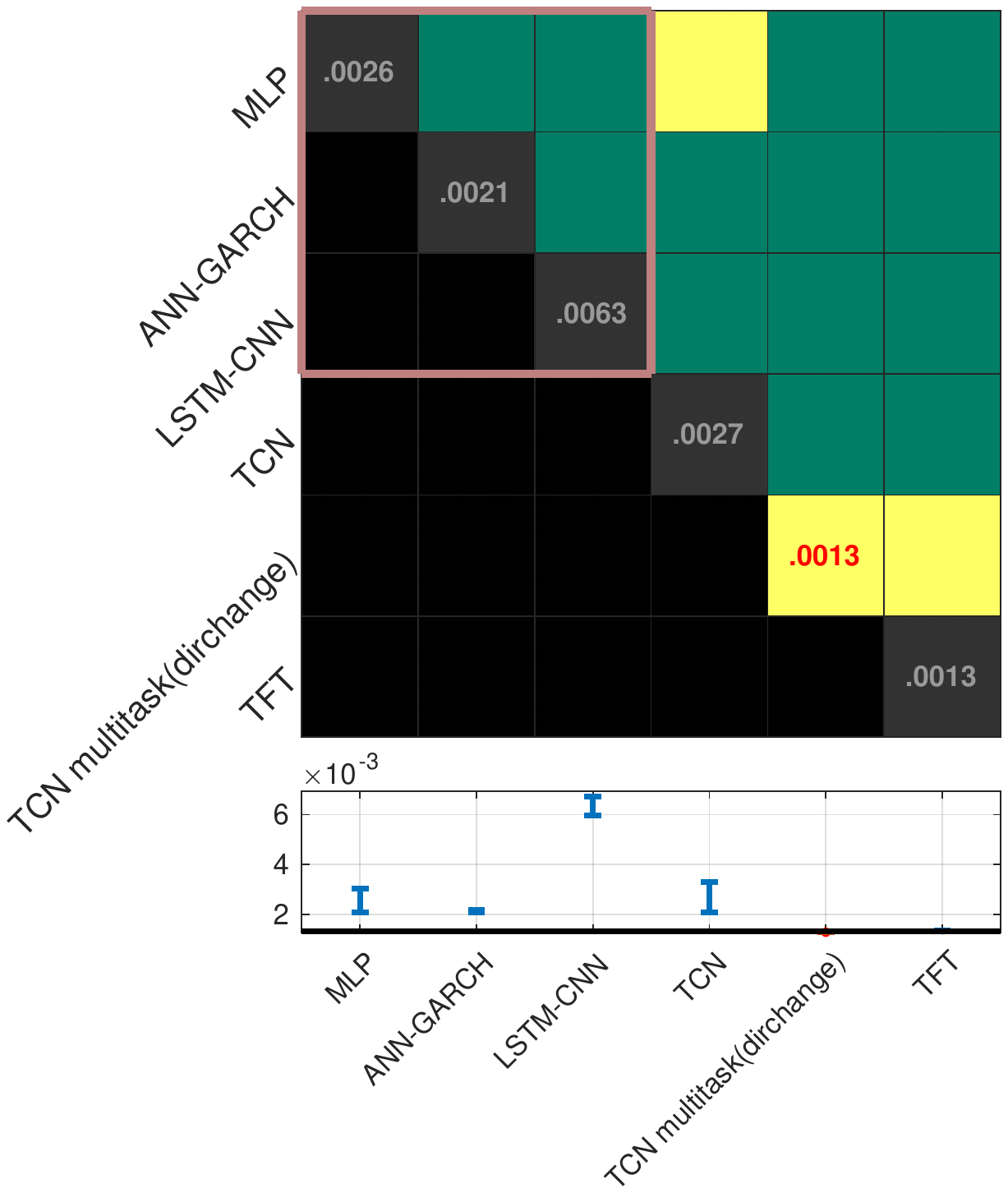}}
\subcaptionbox{Gold}
{\includegraphics[width=0.32\textwidth, trim={3.25cm 7.25cm 4.5cm 7cm}, clip]{Images/goldIV_.pdf}}
\\
\subcaptionbox{Silver}
{\includegraphics[width=0.32\textwidth, trim={3.25cm 7.25cm 4.5cm 7cm}, clip]{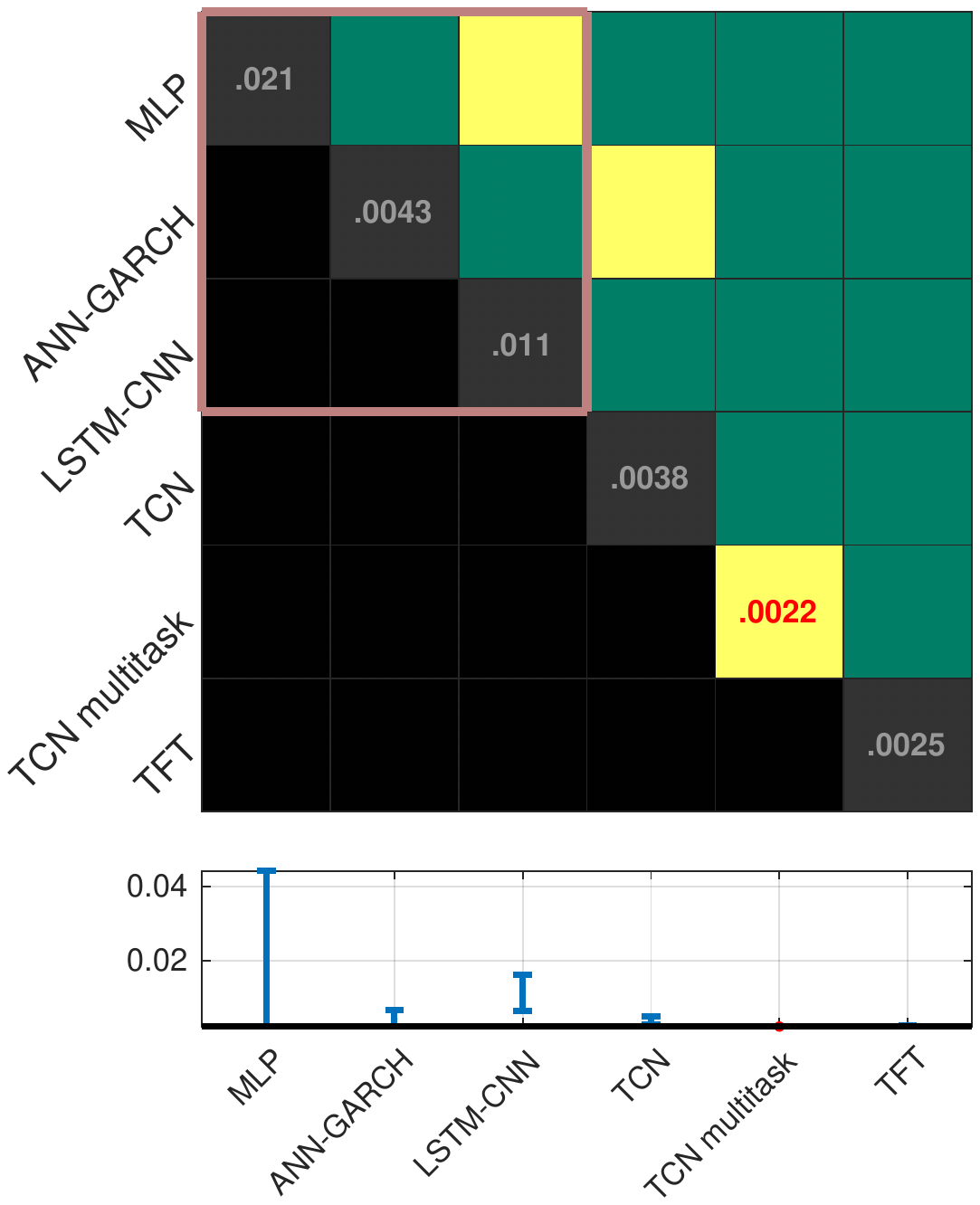}}
\subcaptionbox{Nasdaq}
{\includegraphics[width=0.32\textwidth, trim={2.25cm 6.25cm 4.5cm 7cm}, clip]{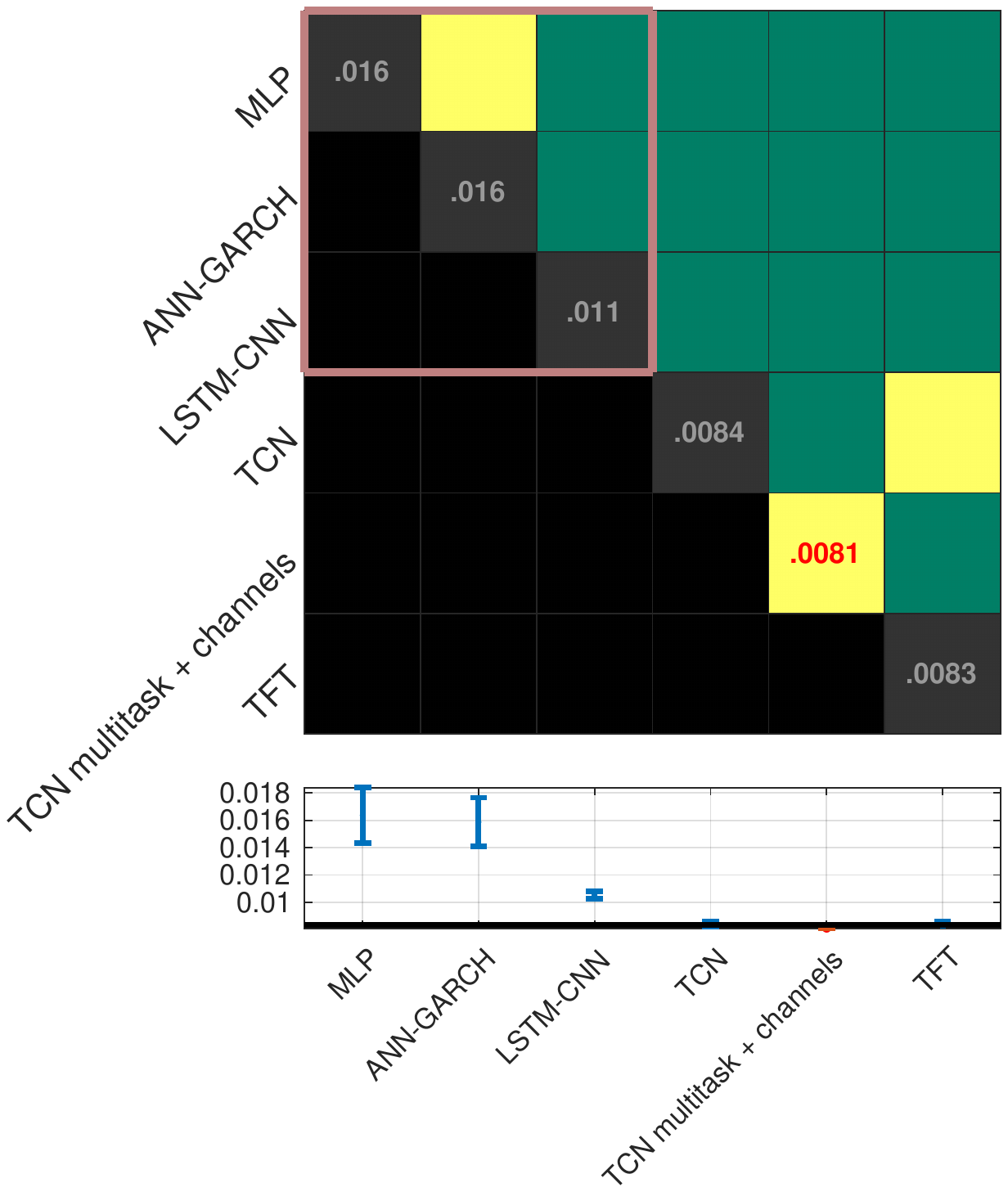}}

\caption{Model comparison of \textbf{implied volatility prediction} using MSE obtained for ten random seed values.}\label{fig:statistical_comparison_IV}
\end{figure*}

Statistical testing results were compiled into ten charts in figures (\ref{fig:statistical_comparison_HV}, \ref{fig:statistical_comparison_IV}), one for each combination of an asset (S\&P500, NDX, gold, silver, and oil) and task (HV and IV forecasting). Each chart  is comprised of a horizontal pane and a 6$\times$6 grid. The horizontal pane shows the distribution of errors from a given model over ten repetitions, indicated by the red and blue vertical lines, relative to the na\"ive forecast, the black horizontal line.

The brackets shown in blue in figures (\ref{fig:statistical_comparison_HV}, \ref{fig:statistical_comparison_IV}) denote that a statistically significant difference could not be found between the corresponding model and the na\"ive model, whilst the brackets in red indicate that a statistically significant difference was found. The 6$\times$6 grid implies that a statistical difference was detected between different models on the $x$ and $y$ axes, with the benchmark models encased in a pink square, and the experimental models outside the pink square. The diagonal of this grid also contains text that shows the average MSE of ten repetitions, with the smallest error indicated in red text. Grey cells indicate that the model either was not found to be statistically different from the na\"ive model, or is worse than it. Yellow cells indicate that no statistical difference was detected between models. They appear on the diagonal as comparing the same models always leads to no difference being found, and in the upper triangle, indicating that there was no statistical difference between the two corresponding models. Green cells indicate a statistically significant difference is detected. They only appear in the upper triangle and indicate that the two corresponding models are different, implying the model with a smaller error is better than the other. The lower triangle of the grid is muted with black cells as the results are mirrored in the upper triangle. Thus, from this set of charts, we can determine if the model with the smallest error forecast (indicated in red text on the diagonal) performs similarly (no statistical difference detected, that is, yellow cells) or better (statistical difference detected, that is, green cells) than other models.


\begin{table*}[t]
\caption{\label{table: statistical significance} Smallest error models (in bold), and models for which no statistically significant difference could be found from the smallest error models.}
\centering
\begin{tabular}{p{0.5cm} p{1.9cm} p{1.25cm} p{1.25cm} p{2.1cm} p{1.25cm} p{1.25cm}}
\toprule
 \multirow{1}{2em}{\textbf{Task}} & \textbf{Model} & \textbf{S\&P500} &\textbf{NDX} & \textbf{Gold} & \textbf{Silver} & \textbf{Oil} \\
\midrule
 \multirow{2}{2em}{HV} & Smallest error & \textbf{TFT}  &   \textbf{TFT}  &   \textbf{ANN-GARCH}  &   \textbf{TCN var.}  &   \textbf{TFT}  \\
 & Equivalent & - & - & TFT  &  TCN  & - \\
\midrule
 \multirow{2}{2em}{IV} & Smallest error & \textbf{TCN var.}  &   \textbf{TCN var.}  &   \textbf{TCN var.}  &   \textbf{TCN var.}  &   \textbf{TCN var.}  \\
 & Equivalent & - & - & - & - & - \\
\bottomrule
\end{tabular}
\end{table*}

\section{Discussion} \label{sec: discussions}

The evaluation results revealed that the experimental models performed better than the benchmark ones, across both defined forecasting tasks, and across all five assets. Statistical significance testing confirmed this, indicating that for almost all combinations of assets and tasks, the best performance belonged to an experimental model, with all of the benchmark models different (and worse) in performance on the basis of statistical significance. The only exception to this was the HV forecasting of gold, for which a statistically significant difference could not be found for one benchmark model (ANN-GARCH) and the experimental models. 

As for why the experimental models perform better than the intelligent models previously used, it is likely a combination of several different reasons. One difference common amongst the experimental models is that they are deeper, that is, they have significantly more neurons. This means that they have more learning capacity, can decode more complex information, and process larger amounts of data. However, this does not mean that simply increasing the number of neurons in any NN-based model would yield the same performance improvements. This is because structure plays an important role. 

The structure of the TCN is similar to that of the CNN and thus shares many of the same benefits. Translational invariance allows the model to detect the same features regardless of where in time it occurs. Weight sharing and a progressively increasing dilation imply that each neuron in the TCN is more efficient than the neurons in an MLP, allowing the TCN to have a wider receptive field and do more processing with the same number of neurons. Hierarchical layering along with residual connections allows the TCN to avoid several problems common to RNNs and LSTMs, such as vanishing gradients and poor long-term memory. Furthermore, the modifications of the TCN variants also play an important role. Forecasting a residual rather than the raw value limits the non-stationarity of the data and encodes the na\"ive forecast directly into the inputs, providing an additional source of information and a good starting point. 

The TFT also differs from previously used NN-based models in many ways. One key difference is the use of multi-headed self-attention, allowing the model to focus on different periods of time conditional on the current value. As an example, this means that if current volatility is low, the model might pay attention to long-term history, whilst if current volatility is high, the model might pay attention to short-term history. Other models like the MLP and TCN can only interpret dependencies from fixed relative time steps, akin to a fixed attention mask. Additionally, the TFT makes better use of certain context variables like cyclic temporal variables (such as the day of the week or month of the year) and market variables (such as the strength of the US dollar, and Federal Reserve credit rates), allowing the model to be better conditioned when making a forecast. The gating mechanism of the TFT also allows for depth in certain areas of the model where it is needed, and shallowness where it is not. Previous models have predefined depth and neuron count, potentially not having enough capacity to learn certain relationships, or being overparameterized and simply remembering training examples rather than generalizing.


These results, however, come with some limitations. The first of which was that the implementation of several models (GARCH and TFT) was open source and thus not necessarily under the same strict control as the other models used. Second, because the purpose of this investigation was to compare different models, rather than to find the best model, the hyperparameter search space may be limited. This does not mean to say that the search space was small, but rather that there likely exists another combination of hyperparameters that would yield better results.

As it seems that DL may offer better volatility forecasting performance than traditional methods, however further research is needed to confirm whether this is true for other models and financial quantities. For instance, there are already several time series adaptations of the Transformer \cite{huang_music_2018, li_enhancing_2019, wu_deep_2020}. The most recent is Autoformer \cite{Wu_2021} which is an extension of Performer, another transformer-based model.  Autoformer, achieves SOTA, with a 38\% relative improvement on six benchmarks, covering five practical applications: energy, traffic, economics, weather and disease. For a survey of Autofromer follow-up models see \cite{Zhu_2023}.
Another class of models is graph neural networks, which recently have been applied to the volatility prediction task \cite{Zhang_2023}. Finally, the use of stochastic NNs \cite{adorf_discrete_1990} or Gaussian process NNs \cite{lee_deep_2018} may have potential as they handle uncertainty well, one of the defining features of volatility. 

Another avenue to enhance the models is to make use of multi-modal data, that is, extend to multiple modes, or types, of data. Each type of data may capture different information, and when combined, could produce better forecasts. Examples of extensions to text exist (e.g., by using Twitter posts \cite{kordonis_stock_2016, oliveira_impact_2017} or news articles \cite{robertson_news_2007, malik_estimating_2011}). Combined with the larger capacity of deeper models, this approach may yield promising results. Furthermore, the use of different paradigms, such as continual learning, could yield even more benefits, especially in the arena of applied volatility forecasting for real-world use \cite{shin_continual_2017}. Further future works could be paved by specifically designing models for financial volatility forecasting, explicitly defined to take advantage of different properties in financial time series and data.

\section{Conclusions} \label{sec: conclusions}

In closing, the DL-based methods have much to offer for financial volatility forecasting. This study implemented several deep learning-based and traditional models and compared these to several experimental recently proposed models for time series prediction. The results indicate that variants of TCN and TFT were able to forecast volatility with less error across almost all assets and tasks investigated, with statistical significance.

\section*{Competing interest}
Although this paper is a research outcome related to a PhD student internship, the authors wish to declare that Euler Capital is a multi-manager platform investing capital across a broad range of trading strategies, focusing on investment, research, and technology by deploying quantitative, systematic, and behavioural-based trading models to transform volatility into opportunity.
The authors have no other conflict of interest to declare.


\begin{acks}
This work was supported by Euler Capital Pty Ltd, under APR Intern Agreement (INT - 0804). We would also like to thank the APR.Intern and \textit{Australian National University} (ANU) for their support to this research through the Australian Postgraduate Research Internship Program (APR.Intern) program. We also gratefully acknowledge funding from the \textit{Australian Government Research Training Program} (AGRTP) Domestic Scholarship for the first author’s PhD~studies. 
\end{acks}

\bibliographystyle{ACM-Reference-Format}
\bibliography{references,sample-base}

\begin{appendices}



\clearpage

\section{Model grid search and input details}\label{appendix: gridsearch and inputs}
\end{appendices}
This section outlines the details of the grid search used for hyperparameter optimisation of the models used, including the list of input variables (\ref{appendix: Table for gridsearch}). 


All benchmark models used grid search as the sole method for tuning hyperparameters \cite{lavalle_relationship_2004, bishop_pattern_2006}, as this was the standard in the literature. This evaluated every combination of hyperparameters within a defined search space against the validation set, that is, trained on the training set, and assessed with the validation set. The hyperparameter combination that performed the best was moved on to be evaluated against the test set, that is, trained on the union of training and validation set, and assessed with the test set. The unmodified TCN was tested with both grid search and BOHB to tune its set of hyperparameters, each given approximately the same wall time. The best model from each method was assessed against the test set, and the search method that performed best overall was used for the remaining TCN experimentation. This strategy could not be applied to the TFT as training was computationally expensive due to the size of the network. Thus, grid search was used in conjunction with the Optuna hyperparameter search method.

\clearpage
\onecolumn

\begin{longtable}[c]{ @{} m{3.5em} r m{32em} @{}}
\caption{Full grid search values for forecasting models}\\

 \toprule
 Model & Hyperparameter & Values \\
 \midrule
 \endfirsthead

 \toprule
 Model & Hyperparameter & Values \\
 \midrule
 \endhead

 \midrule
 \multicolumn{3}{r}{\footnotesize\itshape Continued on the next page}
 \endfoot

 \bottomrule
 \multicolumn{3}{m{\linewidth}}{\footnotesize{\textsuperscript{*}The values for these hyperparameters indicate the lower and upper boundaries of a range, rather then the discrete values, as the hyperparameter optimisation for TFT uses Optuna.}}
 \endlastfoot

 \multirow{2}{3.5em}{GARCH} & Variant & ARCH, GARCH, EGARCH, GJR-GARCH, GARCH with Student's t-distribution \\
 \addlinespace
 & Lag order & 1, 2\\
 \addlinespace
 
 \midrule
 
 \multirow{6}{3.5em}{NAR-MLP} & Lag order & 1, 3, 5, 10, 15, 21\\
 & Number of hidden layers & 1, 2, 4\\ 
 & Neurons per hidden layer & 3, 5, 10, 21 \\ 
 & Activation function & ReLU, Sigmoid, TanH\\ 
 & Learning rate & 0.002, 0.02 \\ 
 & Preprocessing method & Standard scaler, min-max scaler\\
 \addlinespace
 
 \midrule
 
 \multirow{7}{3.5em}{NARX-MLP} & Lag order & 1, 3, 10, 15, 21\\
 & Number of hidden layers & 1, 2, 3, 4\\ 
 & Neurons per hidden layer & 5, 10, 20, 40, 100 \\ 
 & Activation function & ReLU, Sigmoid, TanH\\ 
 & Learning rate & 0.002, 0.02 \\ 
 & Preprocessing method & Standard scaler, min-max scaler\\
 & Additional input variables & See Table \ref{appendix: table of model inputs}\\
 \addlinespace

 \midrule
 
 \multirow{8}{3.5em}{ANN-GARCH}
 & Variant (GARCH) & GARCH, FIGARCH, GJR-GARCH, TARCH, GARCH with Student's t-distribution\\
 & Lag order (GARCH) & 1, 2\\
 & Lag order (ANN) & 3, 8, 21\\
 & Number of hidden layers & 1, 2, 3\\
 & Neurons per hidden layer & 4, 5, 10, 20, 30\\
 & Activation functions & ReLU, Sigmoid\\
 & Preprocessing method & Standard scaler, min-max scaler\\
 & Additional input variables & See Table \ref{appendix: table of model inputs}\\
 \addlinespace
 
 \midrule
 
 \multirow{3}{3.5em}{CNN-LSTM} & Learning rate & 0.0001, 0.001\\
 & Preprocessing method & Standard scaler, min-max scaler\\
 & Receptive field (LSTM) & 1, 5, 10, 21, 42\\
 
 
%
 \multirow{11}{3.5em}{TCN}
 & Residual definition & Difference, or log difference\\
 & Additional tasks & Predict direction, or predict change in direction\\
 & Additional task weight & 0, 0.25, 0.5, 0.75\\
 & Number of filters & 2, 4, 5, 7\\
 & Channels per filter & 20, 25, 30, 35, 40\\
 & Kernel size & 2, 5, 7 \\
 & Dropout & 0.1, 0.3, 0.5 \\
 & Gradient clipping threshold & 0.2, -1 (off) \\
 & Learning rate & 0.001, 0.0001 \\
 & Preprocessing method & Standard scaler, min-max scaler\\
 & Additional input variables & See Table \ref{appendix: table of model inputs}\\
 \addlinespace


 \midrule

 & Gradient clipping threshold\textsuperscript{*} & 0.1, 0.8\\
 & Hidden size\textsuperscript{*} & 16, 128\\
 & Hidden continuous size\textsuperscript{*} & 16, 128\\
 & Number of attention heads\textsuperscript{*} & 1, 8\\
 TFT & Learning rate\textsuperscript{*} & 0.005, 0.5\\
 & Dropout\textsuperscript{*} & 0.1, 0.5\\
 & Encoder length & 10, 21, 126\\
 & Prediction length & 1, 5\\
 & Additional input variables & See Table \ref{appendix: table of model inputs}\\
 
 \label{appendix: Table for gridsearch}
 \end{longtable}

\clearpage





\end{document}